\documentstyle[times,afd-gen,afd-orm,asymetrix,url,pq,a4wide]{article}
\title{Interactive Query Formulation using 
       Point to Point Queries\EOL
       {\normalsize Asymetrix Report 94-1}
}
\author{
   H.A. Proper\\
   Asymetrix Research Laboratory\\
   Department of Computer Science\\
   University of Queensland\\
   Australia 4072\\
   E.Proper@acm.org
}
\date{\Version}

   \input{epsf}
   \def\Figurepath{./}
   \def\Scale{0.9}
   \def\epsfsize#./##2{\Scale#./}

\begin{document}
   \maketitle
   {\sc Published as:}
\begin{quote}
  H.A.~(Erik) {Proper}. {Interactive Query Formulation using Point to Point Queries}. Technical report, Asymetrix Research Laboratory, University of Queensland, Brisbane, Queensland, Australia, 1994.
\end{quote}

   \begin{abstract}
   Effective information disclosure in the context of databases with a large conceptual schema is known to 
   be a non-trivial problem. In particular the formulation of ad-hoc queries is a major problem in such
   contexts. Existing approaches for tackling this problem include graphical query interfaces, query by
   navigation, and query by construction. In this article we propose the {\em point to point query 
   mechanism} that can be combined with the existing mechanism into an unprecedented computer supported 
   query formulation mechanism.

   In a point to point query a path through the information structure is build. This path can then be used to 
   formulate more complex queries. A point to point query is typically useful when users know some object 
   types which are relevant for their information need, but do not (yet) know how they are related in the 
   conceptual schema. Part of the point to point query mechanism is therefore the selection of the most 
   appropriate path between object types (points) in the conceptual schema.

   This article both discusses some of the pragmatic issues involved in the point to point query mechanism, 
   and the theoretical issues involved in finding the relevant paths between selected object types.
\end{abstract}

   \section{Introduction}
Most present day organisations make use of some automated information system. This usually means that 
a large body of vital corporate information is stored in these information systems. As a result an obvious, 
yet crucial, function of information systems is the support of disclosure of this information. Without a 
set of adequate information disclosure avenues an information system becomes worthless since there is no 
use in storing information that will never be retrieved. An adequate support for information disclosure, 
however, is far from a trivial problem. Most query languages do not provide any support for the users in 
their quest for information. Furthermore, the conceptual schemata of real-life applications tend to be quite 
large and complicated. As a result, the users may easily become 'lost in conceptual space' and they will 
end up retrieving irrelevant (or even wrong) objects and may miss out on relevant objects. Retrieving 
irrelevant objects leads to a low {\em precision}, missing relevant objects has a negative impact on the 
{\em recall} (\cite{Book:83:Salton:IntroIR}).

The disclosure of information stored in an information system has some clear parallels to the disclosure 
problems encountered in {\em document retrieval systems}. To draw this parallel in more detail, we quote 
the information retrieval paradigm as introduced in \cite{Report:91:Bruza:StratHypmed}. The paradigm 
starts with an individual or company having an {\em information need} they wish to fulfil. This need is 
typically a vague notion and needs to be made more concrete in terms of an {\em information request} 
(the query) in some (formal) language. The information request should be as good as possible a 
description of the information need. The information request is then passed on to an automated system, or 
a human intermediary, who will then try to fulfil the information request using the information stored in 
the system. This is illustrated in the {\em information disclosure}, or {\em information retrieval 
paradigm}, presented in \SRef{\IRParadigm} which is taken from \cite{Report:91:Bruza:StratHypmed}. 
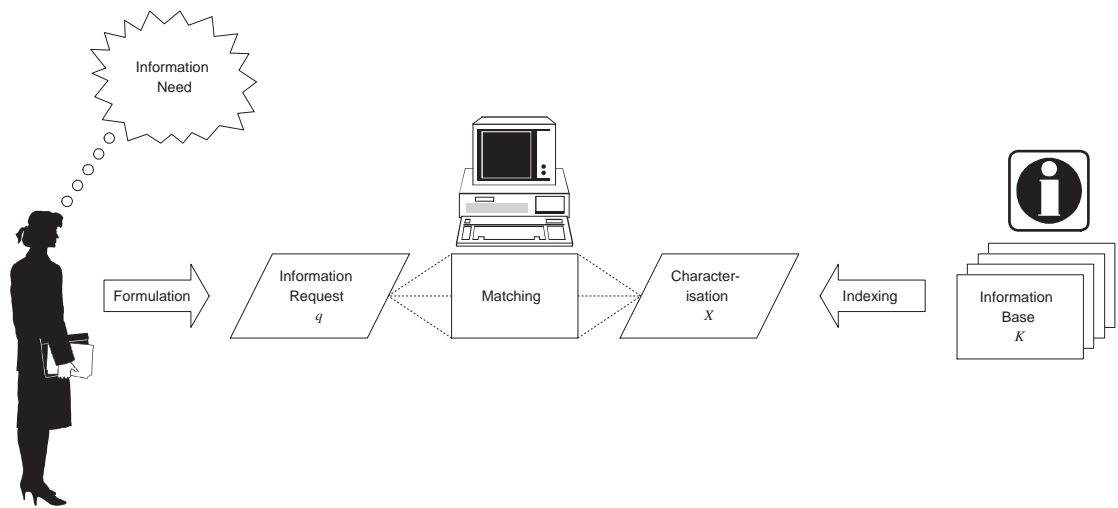
{\def\Scale{0.50} \EpsfFig[\IRParadigm]{The information retrieval paradigm}}

We now briefly discuss why the information retrieval paradigm for document retrieval systems is also 
applicable for information systems. For a more elaborate discussion on the relation between information 
systems and document (information) retrieval systems in the context of the information retrieval 
paradigm, refer to \cite{PhdThesis:94:Proper:EvolvConcModels}. In the paradigm, the retrievable 
information is modelled as a set \Cal{K} of {\em information objects} constituting the {\em information 
base} (or population).

In a document retrieval system the information base will be a set of documents 
(\cite{Book:83:Salton:IntroIR}), while in the case of an information system the information base will 
contain a set of facts conforming to a conceptual schema. Each information object $o \in \Cal{K}$ is 
{\em characterised} by a set of descriptors $\chi(o)$ that facilitates its disclosure. The characterisation of 
information objects is carried out by a process referred to as indexing. In an information system, the stored 
objects (the population or information base) can always be identified by a set of (denotable) values, the 
identification of the object. For example, an address may be identified as a city name, street name, and 
house number. The characterisation of objects in an information system is directly provided by the 
reference schemes of the object types. 

The actual information disclosure is driven by a process referred to as {\em matching}. In document 
retrieval applications this matching process tends to be rather complex. The characterisation of documents 
is known to be a hard problem (\cite{Article:77:Maron:InfRetr}, \cite{Book:86:Craven:SringIndexing}), 
although newly developed approaches turn out to be quite successful 
(\cite{Book:89:Salton:AutomTextProc}). In information systems the matching process is less complex as 
the objects in the information base have a more clear characterisation (the identification). In this case, the 
identification of the objects (facts) is simply related to the query formulation $q$ by some (formal) query 
language.

The remaining problem is the query formulation process itself. An easy and intuitive way to formulate 
queries is absolutely essential for an adequate information disclosure. Quite often, the quest from users to 
fulfil their information need can be aptly described by (\cite{PhdThesis:92:Bruza:IRHypmed}): 
\begin{quote} \it
   I don't know what I'm looking for, but I'll know when I find it.
\end{quote}

In document retrieval systems this problem is attacked by using {\em query by navigation} 
(\cite{Report:91:Bruza:StratHypmed}, \cite{PhdThesis:92:Bruza:IRHypmed}) and {\em relevance 
feedback} mechanisms (\cite{Article:89:Rijsbergen:IRLogic}). The query by navigation interaction 
mechanism between a searcher and the system is well-known from the Information Retrieval field, and 
has proven to be useful. It shall come as no surprise that these mechanisms also apply to the query 
formulation problem for information systems. In \cite{Report:92:Burgers:PSMIR}, 
\cite{Report:93:Burgers:PSMIR}, \cite{Report:93:Hofstede:DisclSupport}, 
\cite{PhdThesis:94:Proper:EvolvConcModels}  such applications of the {\em query by navigation} and 
{\em relevance feedback} mechanisms have been described before. When combining the query by 
navigation and manipulation mechanisms with the ideas behind visual interfaces for query formulation as 
described in e.g. \cite{Article:92:Auddino:SUPERVisual}  and \cite{Article:94:Rosengren:VisualER} 
powerfull and intuitive tools for computer supported query formulation become feasible. Such tools will 
also heavily rely on the ideas of direct manipulation interfaces 
(\cite{Article:83:Shneiderman:DirectManip}) as used in present day computer interfaces. 

One important step in the improvement of the information disclosure of information 
systems, is the introduction of query languages on a conceptual level. 
Examples of such conceptual query languages are 
RIDL (\cite{Report:82:Meersman:RIDL}), 
LISA-D (\cite{Report:91:Hofstede:LISA-D}, \cite{Report:92:Hofstede:LISA-DPromo}), 
and FORML (\cite{Article:92:Halpin:Subtyping}). 
By letting users formulate queries on a conceptual level, they are safeguarded 
from having to know the exact mapping to internal representations (e.g. a set 
of tables conforming to the relational model), as they would be required when 
formulating queries in a non conceptual language such as SQL. 
The next step is to introduce ways to support users in the formulation of queries in 
such conceptual query languages (CQL). 

In line with the above discussed information retrieval paradigm and the notion of relevance feedback, a 
query formulation process (both for a document retrieval system, and an information system) can be said 
to roughly consist of the following four phases:
\begin{enumerate}
   \item {\em The explorative phase}. What information is there, and what does it mean?
   \item {\em The constructive phase}. Using the results of phase 1, the actual query is formulated.
   \item {\em The feedback phase}. The result from the query formulated in phase 2 may not be
       completely satisfactory. In this case, phases 1 and 2 need to be re-done and the result refined.
   \item {\em The presentation phase}. In most cases, the result of a query needs to be incorporated
       into a report or some other document. This means that the results must be grouped or aggregated
       in some form. 
\end{enumerate}
Depending on the user's knowledge of the system, the importance of the respective phases may change. 
For instance, a user who has a good working knowledge of the structure of the stored information may not 
require an elaborate first phase and would like to proceed with the second phase as soon as possible.
 
In this paper, we discuss one of the mechanisms to support automated disclosure of information stored in information 
systems. As stated before, the related notions of {\em query by navigation} and {\em query by construction} have 
already been discussed in \cite{Report:92:Burgers:PSMIR}, \cite{Report:93:Proper:DisclSch}, 
\cite{PhdThesis:94:Proper:EvolvConcModels}. This article is concerned with the {\em point to point query} (PPQ) 
mechanism as an additional avenue for information disclosure. A point to point query starts by selecting two or more 
object types from a conceptual schema. Then the system will return a list of possible (non cyclic) paths through the 
information structure between the specified object types. For obvious reasons, the paths in this list should be ordered 
according to some relevance criterion. This style of querying corresponds to a situation in which users know some 
aspects (object types) about which they want to be informed, but do not yet know the exact details of their information 
need and the underlying information structure. The query by navigation mechanism, on the other hand, is intended to 
support users who do not have an overview of the stored information.

Dispite how simple the above scenario may seem, the point to point query mechanism 
required to realise this query is far from trivial. There are two 
main problems involved. Firstly, all (non-cyclic) paths through the conceptual schema must be found. 
This corresponds to finding all (non-cyclic) paths between two nodes in a graph,
which is in general an exponential (NP hard) problem. 
(Finding the shortest paths is polynomal!) 
The second main problem is the order in which the results should be 
presented to the user. It is clear that (especially when there is an abundance of possible paths to choose 
from) the alternatives should be presented to the user in some order of relevance. We believe to have 
found an approach to these two problems that makes a point to point query mechanism feasible.

The structure of this article is as follows. 
In \SRef{PPQEx}, we discuss an example PPQ session, and elaborate 
briefly on its integration with query by navigation and query by 
construction. 
Section \ref{CSGraph} deals with the representation of a conceptual schema as a graph. 
Searching for a path through this graph is covered in \SRef{Search}. 
In \SRef{Reduce} an optimisation strategy is introduced based on a 
pre-compilation of the conceptual schema graph. Finally, \SRef{Concl} 
concludes this article.
For the reader who is unfamiliar with the notation style used in this report,
it is advisable to first read \cite{AsyReport:94:Proper:Formal}.

   \section{An Example PPQ Session}
\SLabel{section}{PPQEx}

In this section we discuss a sample session involving a point to point query, and also discuss briefly 
the relationship to query by navigation and query by construction. The discussed example operates on 
a conceptual schema for the administration of the election of American presidents. The example 
schema itself is not shown; the structure of the domain will become clear from the sample session. 
Note that the quality of the verbalisations of paths expressions used in the 
examples in this section should be improved.
However, this is the subject of further research. 

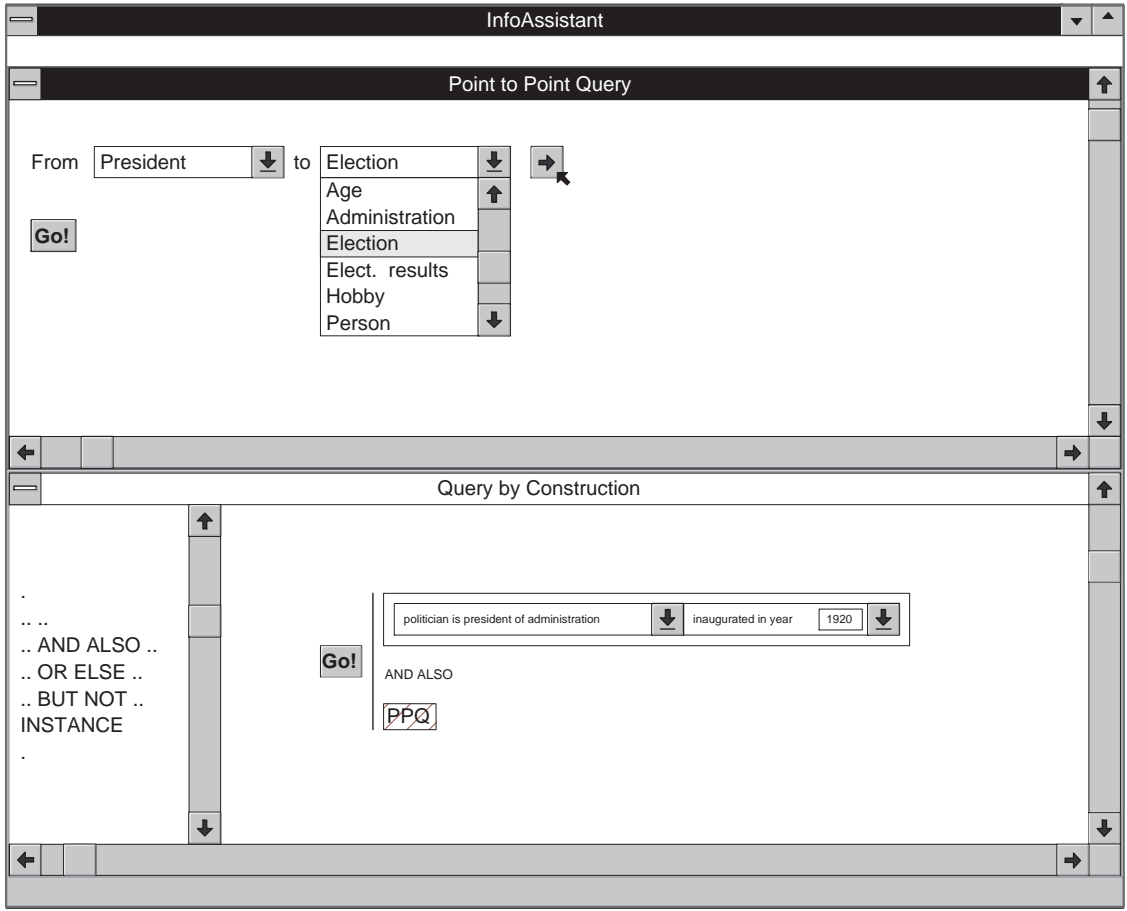
{\def\Scale{0.70} \EpsfFig[\PPQStart]{Building a PPQ query}}
In \SRef{\PPQStart}, a possible screen is depicted for building queries using a point to point query 
mechanism. The upper window is concerned with the point to point query itself, whereas the lower 
window contains the complete query under construction. When specifying a point to point query a 
user specifies a sequence of object types: the points. For each point, the user is offered a listbox 
containing all object types present in the conceptual schema. The order of the object types in the listbox should 
preferably be based on some notion of conceptual importance 
(\cite{Article:94:Campbell:Abstraction}). In \SRef{\PPQExtension} an existing point to point query 
path from president to election is extended with another point.
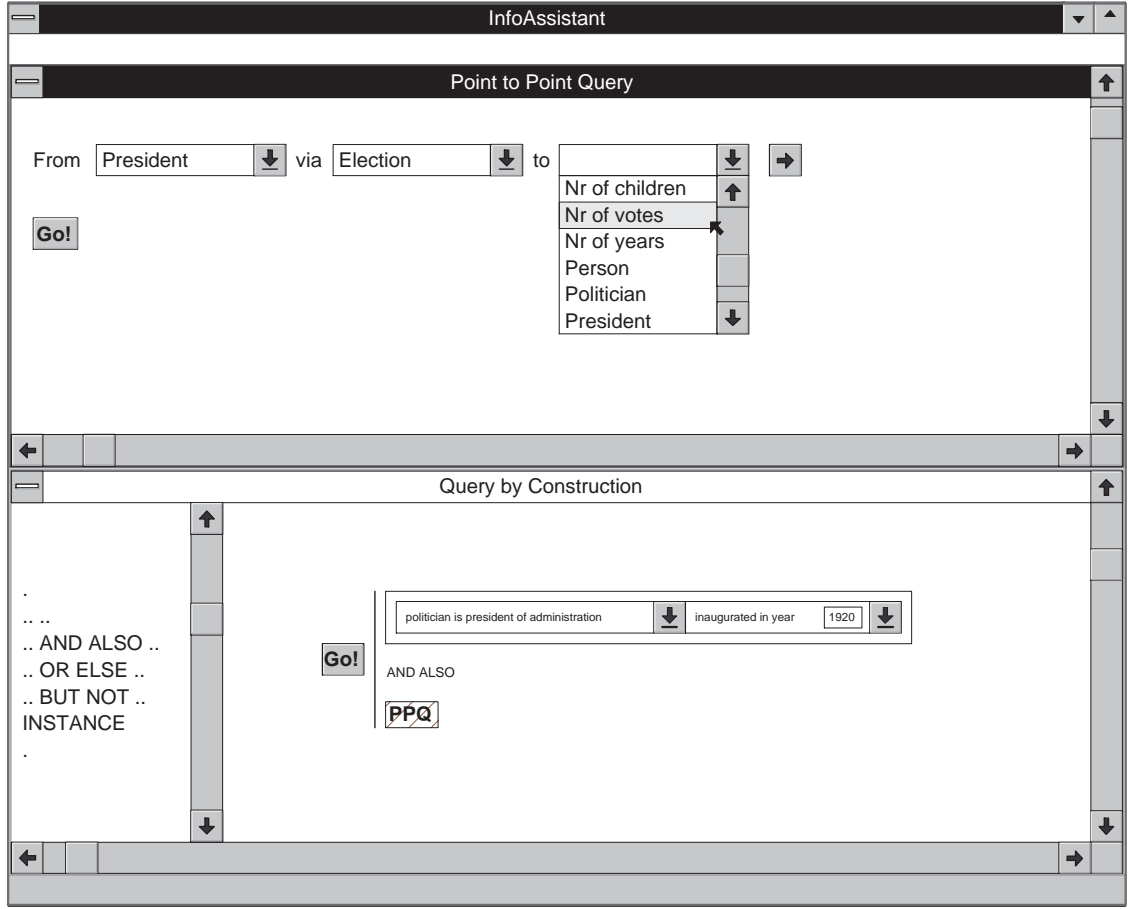
{\def\Scale{0.70} \EpsfFig[\PPQExtension]{Extending the PPQ path}} 

An important underlying practical issue is whether the selection of the points in a point to point query 
should be done graphically or textually. The theoretical discussions in the remainder of this article are not 
influenced by such a choice; but this should be taken into serious consideration when implementing the 
point to point query mechanism. Although the ideal situation may seem to be a graphical selection 
mechanism using the conceptual schema itself by clicking on object types ellipses, 
this may turn out to be impossible due to the limited size of PC screens. Graphical 
based visual query formulation interfaces (\cite{Article:92:Auddino:SUPERVisual}, 
\cite{Article:94:Rosengren:VisualER}) work well for small schemes or queries covering only a small 
sub schema. However, when a large conceptual schema is involved graphical languages may turn out 
to be to space consuming.
{\def\Scale{0.70} \EpsfFig[\PPQComplete]{Completing a PPQ}}

After all points of the point to point query have been specified, the point 
to point query can be transformed into a proper query (i.e. a path through 
the conceptual schema) by pressing the \SF{Go!} button in the point to point 
query window. 
In \SRef{\PPQComplete}, this process is illustrated. 
The sample PPQ involves three points. 
Therefore, two paths through the conceptual schema will result. 
We now shift our attention from the point to point query window to the query 
by construction window. 
Note that the small box containing the \SF{PPQ} abbreviation is now replaced 
by the paths resulting from the point to point query (i.e. \SF{President 
winning election which resulted in nr of votes}). 
The system initially inserts a most likely path. 
The user can, however, select alternative paths using a listbox. 
Note that not all alternative paths between the two points are listed in 
the listbox. 
The reason for this is the NP completeness of the path searching problem. 
To avoid the NP completeness problem, only the best paths are listed 
initially. 
However, potentially all paths can be selected (which still remains NP 
complete) by repeatedly selecting the \SF{MORE} option. 
In the remainder of this article we will discuss this in more detail.

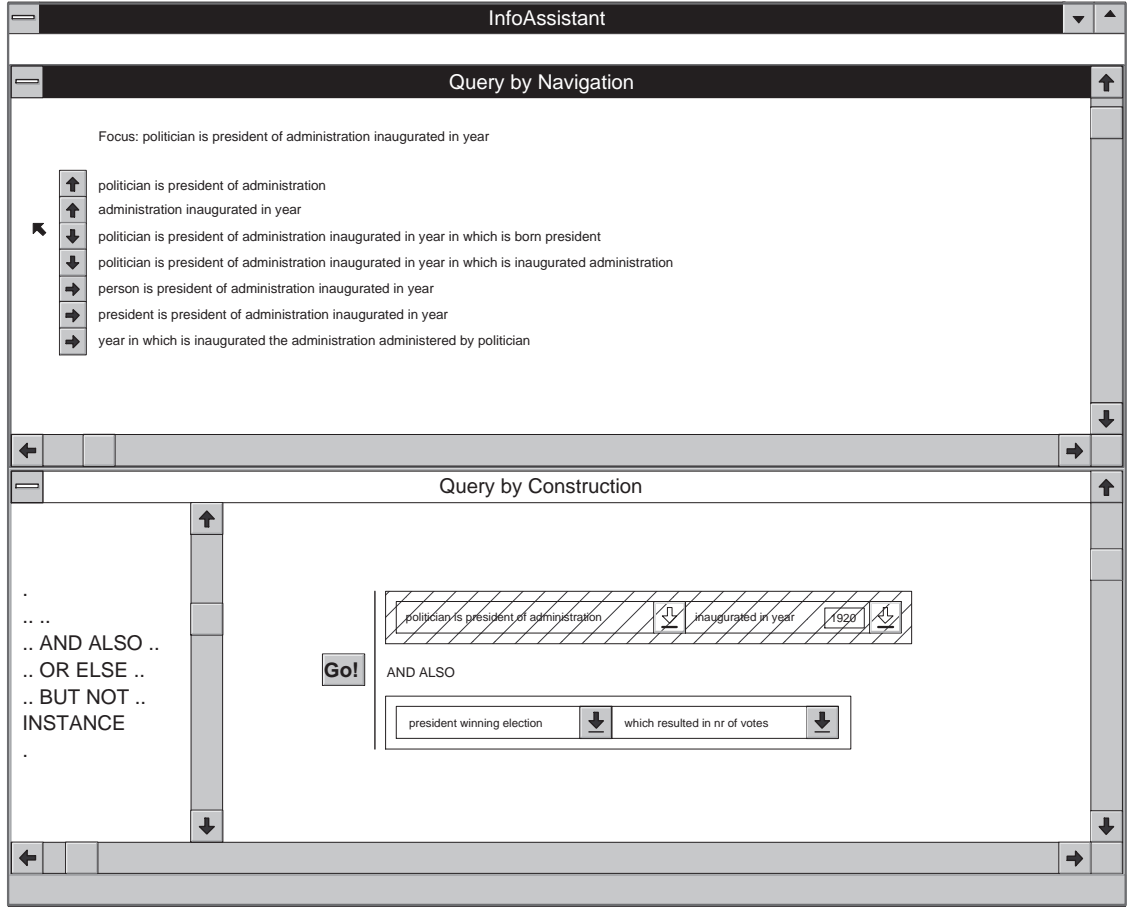
{\def\Scale{0.70} \EpsfFig[\PPQNavigate]{Switching to query by navigation}}
Since every path resulting from a query by navigation session connects two 
points in the conceptual schema, any path through the conceptual schema 
displayed in the query by construction screen can be used as a starting point
for a query by navigation session, and vice versa. 
This is illustrated in \SRef{\PPQNavigate}. 
In this session, the user has selected the box which contains the two paths 
\SF{politician is president of administration} and \SF{inaugurated in year} 
for a query by navigation session. 
The upper window now displays a node in the query by navigation session, with 
the path \SF{politician is president of administration inaugurated in year}
as its focus. 
If the user had selected the \SF{inaugurated in year} listbox, the 
initial focus would have been \SF{administration inaugurated in year}.

The query by construction window is basically a syntax directed editor. In the left part of the window 
all possible constructs from the query language are listed. In our examples we have used the constructs 
defined in LISA-D. Once the FORML and LISA-D languages have been merged, a more complete 
language for the query by construction part will result.

   \section{A Conceptual Schema as a Graph}
\SLabel{section}{CSGraph}

For the purpose of finding a path between object types in a conceptual schema, the schema first needs 
to be translated to a graph. We start out from a formalisation of ORM based on the one used in 
(\cite{Report:94:Halpin:ORMPoly}). However, since only a very limited part of the formalisation is 
needed, we do not cover the formalisation in full detail.

A conceptual schema is presumed to consist of a set of types $\Types$. 
Within this set of types two subsets can be distinguished: 
the relationship types $\RelTypes$, and the object types $\ObjTypes$. 
Furthermore, let $\Preds$ be the set of roles in the conceptual schema. 
The fabric of the conceptual schema is then captured by two functions 
and two predicates. 
The set of roles associated to a relationship type is provided by the 
partition: $\Roles: \RelTypes \Func \Powerset(\Preds)$. 
Using this partition, we can define the function $\Rel$ which returns for each
role the relationship type in which it is involved:
  $\Rel(r) = f \iff r \in \Roles(f)$.
Every role has an object type at its base called the player of the role, 
which is provided by the function: $\Player: \Preds \Func \Types$. 
Subtyping and polymorphism of object types is captured by the predicates 
$\Spec \subseteq \ObjTypes \Carth \ObjTypes$ and 
$\Poly \subseteq \ObjTypes \Carth \ObjTypes$ respectively. 
For any ORM conceptual schema the following (undirected) labelled graph 
$G = \tuple{N,E}$ can be defined:
\begin{eqnarray}
   N  &\Eq& \Types\\
   E  &\Eq&    \Set{\tuple{\setje{\Player(r), \Rel(r)}, r}}{r \in \Preds}\\
      &\union& \Set{\tuple{\setje{x,y},\Spec}}{x \Spec y}\\
      &\union& \Set{\tuple{\setje{x,y},\Poly}}{x \Poly y}
\end{eqnarray}
The edges in the resulting graph have the format $\tuple{\setje{x,y},l }$, 
where $x$ and $y$ are the source/destination (no order) of the edge, and 
$l$ is the label of the edge. 
The labels on the edges either result from the roles in the relationship types
(2), or they result from specialisation and polymorphism (3,4).
In the remainder, the graph $G$ will be used as an implicit parameter for all
introduced functions and operations. 
As a convention, the nodes of graph $G$ are accessed by $G.N$, and the 
edges by $G.E$.

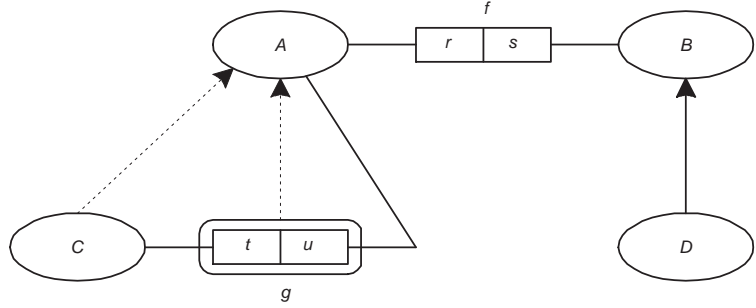
\EpsfFig[\ExampleCS]{Example Conceptual Schema}
As an example consider the conceptual schema depicted in \SRef{\ExampleCS}. For this schema we 
have:
\[ \begin{array}{lllcl}
   \Types     &=& \setje{A, B, C, D, f, g} && \Roles(f) = \setje{r,s}, \Roles(g) = \setje{t,u}\Eol
   \RelTypes  &=& \setje{f, g}             && \Player(r) = A, \Player(s) = B, \Player(t ) = C, \Player(u) = A\Eol
   \ObjTypes  &=& \setje{A, B, C, D}	   && A \Poly C, A \Poly g\Eol
   \Preds     &=& \setje{r,s,t,u}          && D \Spec B
\end{array} \]
From this schema the graph as depicted in \SRef{\ExampleGraph} can be derived.
\EpsfFig[\ExampleGraph]{Example Graph}

For point to point queries paths in the graph need to be found. 
In this article, a path through the graph is denoted as a sequence of 
alternating nodes and labels:
\[  [x_0,l_1,x_1, \ldots,l_n,x_n] \]
Note that if $x$ is a node, then $[x]$ denotes the path consisting of
node $x$ only.
In the remainder $\SeqConc$ is used as the concatenation operation for 
sequences.
Not all alternating sequences of nodes and labels correspond to a proper path. 
For a path to be a proper one, it must adhere to two properties:
\begin{enumerate}
   \item The nodes in the path must originate from the graph:
      $\Al{0 \leq i \leq n}{~x_i \in G.E~}$.
   \item The labels in the path must originate from the proper edges
      in the graph:
      \[ \Al{1 \leq i \leq n}{~\tuple{\setje{x\Sub{i-1},x_i},l_i} \in G.N~}\]
\end{enumerate}
In the remainder of this article, \SF{Path} denotes the set of all valid paths
for any graph $G$ resulting from an ORM schema. 
On such paths the following three functions can be defined: 
   $\SF{Begin}: \SF{Path} \Func G.N$, 
   $\SF{End}: \SF{Path} \Func G.N$, and 
   $\SF{Length}: \SF{Path} \Func \N$ , 
which are identified by:
   \[ \SF{Begin}( [x_0,l_1,x_1, \ldots, l_n,x_n]) ~\Eq~ x_0 \]
   \[ \SF{End}(   [x_0,l_1,x_1, \ldots, l_n,x_n]) ~\Eq~ x_n \]
   \[ \SF{Length}([x_0,l_1,x_1, \ldots, l_n,x_n]) ~\Eq~ n   \]
Furthermore, the $\in$ and $\not\in$ operations can be extended to paths as well, expressing the 
occurrence of a node on a path:
   \[ x \in [x_0,l_1,x_1, \ldots, l_n,x_n] \iff  
      x \in \setje{x_0, \ldots, x_n} \]
   \[ x \not\in p \iff \lnot(x \in p) \]
A {\em badness} level is associated with every path through the conceptual 
schema, expressing its conceptual irrelevance. 
The badness is used to order the alternative paths in the listboxes. 
Badness is defined in terms of a penalty point system where a high penalty point score corresponds to 
a low conceptual relevance. Two ways of earning penalty points exist: the relative conceptual 
irrelevance of the object types in the path, and the length of the path. 

For the first class of penalty points the existence of a function: $\CWeight: \Types \Func \N$ is 
presumed. This function should capture the conceptual importance of the types in the conceptual 
schema, which can for instance be derived from the abstraction level at which the type is present 
(\cite{Article:94:Campbell:Abstraction}). 
For each object type occurring in a path, the number of penalty points added 
to the total badness of the path depends on the deviation of its conceptual 
importance from the maximum conceptual importance in the conceptual schema. 

The second way for a path to earn penalty points is the length of the path. For every object type 
occurring in the path a basic amount of penalty points is added. In order to maintain uniformity of the 
penalty points added, this basic amount is set equal to the maximum conceptual importance of object 
types in the schema. Finally, sometimes one would like to be able to control the influence of the two 
ways to earn penalty points in the final outcome. For this purpose, we introduce the (user definable) 
constant $C_{\it weight } \in [0,1]$. 
This leads to the following definition of the badness of a non-cyclic path 
in a graph $G$:
   \[ \SF{Badness}: \SF{Path}  \Func \N \]
   \[ \SF{Badness}(p) ~\Eq~ 
        \Sigma\Sub{o \in p}
          \left( C_{\it weight} \Carth 
            ({\it MaxCWeight} - \CWeight(o)) + (1 - C_{\it weight}) 
            \times {\it MaxCWeight} \right) \]
where ${\it MaxCWeight} = \Max\Sub{x \in G.N} \CWeight(x)$.
Note the $\in$ operation used in the expression $o \in p$ is the above defined 
inclusion operation for paths through graphs, and that an object type $o$ only
occurs at most once in p and that therefore the summation over $o \in p$ is 
correct with respect to the length of the path. 
An important property of the $\SF{Badness}$ function is the following: 
\begin{lemma}
   The function \SF{Badness} is monotonous strict increasing, i.e.:
   \[ \mbox{if~} 
         p \SeqConc q \in \SF{Path} 
      \mbox{~is an acyclic path and $q$ is non-empty, then~} 
         \SF{Badness}(p) < \SF{Badness}(p \SeqConc q) 
   \]
\end{lemma}
\begin{proof}
   Follows directly from the definition of \SF{Badness} and the observation that the set 
   $\Set{o}{o \in p}$ is a proper subset of $\Set{ o }{ o \in p \SeqConc q}$ 
\end{proof}
The above property allows us to incrementally search for the paths with the lowest badness since the 
badness of a path never decreases when extending it. Note that one might also want to introduce 
additional fitness factors. 
For instance, one could take the correlation of the (verbalisation of the) 
path to a set of keywords describing the user's interests into consideration.
However, the badness should remain a strict monotonous increasing function.

   \section{The Quest}
\SLabel{section}{Search}

This section is concerned with finding a path through the conceptual schema (graph) between two points 
(nodes). Although a point to point query typically involves more then two object types, it can always be 
expressed as a combination of a set of point to point queries over two points. As an example consider 
\SRef{\PPQComplete}. The newly added point to point query involves three points and is represented as 
two point to point queries (listboxes) over two points.

In searching paths between two points (nodes) in the graph, an incremental strategy is followed. Two 
pools of paths are maintained during the entire search: a pool $P$ of paths which could lead to a 
possible solution and a pool $S$ of found solutions. In every step (increment) of the algorithm these pools 
are updated. The best (lowest badness) potential solutions in pool $P$ are selected for further extension. 
By selecting the best paths in $P$ for further extension it can be guaranteed that the first solutions found 
are the ones with the lowest badness. Within a pool of possible solutions $P$, and in the context of a 
graph $G$, the set of best candidates are provided by:
\[ \begin{array}{l}
   \SF{Best}: \Powerset(\SF{Path}) \Func  \Powerset(\SF{Path})\Eol
   \SF{Best}(P) ~\Eq~ 
     \Set{p \in P}{\SF{Badness}(p) = \Min\Sub{q \in P}\SF{Badness}(q)}
\end{array}\]
The first operation we introduce calculates the increment as described above 
for a pair of pools. 
It tries to extend the paths in $P$, and updates the set of found solutions 
in $S$ if new ones have been found. 
For any start node $f$ and end node $t$, we define the increment operator as:
\[ \begin{array}{l}
   \SF{Increment}\Sub{f,t:}: \Powerset(\SF{Path}) \Carth \Powerset(\SF{Path}) \Func 
      \Powerset(\SF{Path}) \Carth \Powerset(\SF{Path}) \Carth \Powerset(\SF{Path})\Eol
   \SF{Increment}\Sub{f,t}(P,S) \Eq  
      \left\{ \begin{array}{ll}
         \tuple{P',S',R'} & \mbox{if~} P \neq \emptyset\Eol
         \tuple{P,S,S}    & \mbox{otherwise}
      \end{array} \right.
\end{array} \]
where:
\begin{eqnarray*}
   N  & = & \Set{
               p \SeqConc [n,l]
            }{
               p \in \SF{Best}(P) \land 
               \tuple{\setje{\SF{End}(p),n},l} \in G.E \land n \not\in p 
            }\\
   S' & = & S \union \Set{s \in N}{\SF{End}(s) = t}\\
   P' & = & (P \union N) \SetMinus \SF{Best}(P) \SetMinus S'\\
   R' & = & \Set{r \in S'}{\SF{Badness}(r) \leq \Min\Sub{q \in P'}\SF{Badness}(q)}
\end{eqnarray*}
For defining the set of (best) extended paths $N$, all best paths ($p \in \SF{Best}(P)$) in 
the existing pool of possible solutions are extended with an appropriate edge from the graph 
($\tuple{\setje{\SF{End}(p),n},l} \in G.E$) while maintaining acyclicity ($n \not\in p$). The new set of 
solutions ($S'$) is simply the old set of solutions extended with the solutions found after extending the 
best paths. In the new pool of possible solutions ($P'$) the newly found solutions are removed since they 
should not be extended any further. Although a path in $S'$ has the proper begin and end point it is not 
considered to be a proper solution until it has a lower badness then the paths in the pool of potential 
solutions $P'$. The set of proper solutions is returned in $R'$.

For the increment operation we have the following property:
\begin{lemma}
   If $\SF{Increment}\Sub{f,t}(P,S)   = \tuple{P',S',R'}$ and 
      $\SF{Increment}\Sub{f,t}(P',S') = \tuple{P'',S'',R''}$, then:
       \[ R' \subseteq R'' \land 
          \Al{r \in R'' \SetMinus R'}{
              \SF{Badness}(r) > \Max\Sub{q \in R'}\SF{Badness}(q)
          } 
       \]
\end{lemma}
\begin{proof}
   We first prove that $R' \subseteq R''$.

   If $r \in R'$, then since $R' \subseteq S' \subseteq S''$ we also have 
   $r \in S''$. Furthermore, from the definition of $R'$ follows: 
   $\SF{Badness}(r) \leq \Min\Sub{q \in P'}\SF{Badness}(q)$. 
   Due to the monotonic behaviour of the \SF{Badness} function,
   we immediately have: 
      $\SF{Badness}(r) \leq \Min\Sub{q  \in P''}\SF{Badness}(q)$, 
   since $P''$ contains the extended paths. 
   From the definition of $R'$ then follows that $r \in R'$.

   Now we prove that 
     $\Al{r \in R'' \SetMinus R'}{
          \SF{Badness}(r) > \Max\Sub{q \in R'}\SF{Badness}(q)
      }$. 
   If $r \in R'' \SetMinus R'$, then $r \not\in R'$. From this and the 
   definition of $R'$ follows that $r \not\in S'$ or 
     $\SF{Badness}(r) > \Min\Sub{q \in P'}\SF{Badness}(q)$. 
   So we have:
   \begin{enumerate}
      \item Let $r \not\in S'$. Since $r \in R'' \SetMinus R'$, we know that $r$ 
         is a newly found solution in $R''$. So there is a $p \in \SF{Best}(P')$ 
         and an $e \in G.E$ such that $p \SeqConc [e] = r$.
         From the monotonicity of \SF{Badness}, it then follows 
           $\SF{Badness}(p) < \SF{Badness}(r)$.
 
         If $x \in R'$, then from the definition of $R'$ follows that 
         $\SF{Badness}(x) \leq \Min\Sub{q \in P'}\SF{Badness}(q)$. 
         Since we just concluded that $\SF{Badness}(p) < \SF{Badness}(r)$ for a 
         certain $p \in P'$, we at least have 
            $\Min\Sub{q \in P'}\SF{Badness}(q) < \SF{Badness}(r)$ 
         Since we also have $\SF{Badness}(x) < \SF{Badness}(r)$,
         we in particular have: 
            $\SF{Badness}(r) > \Max\Sub{q \in R'}\SF{Badness}(q)$.

      \item Let $\SF{Badness}(r) > \Min\Sub{q \in P}\SF{Badness}(q)$. 
         From the definition of $R'$ follows that if $x \in R'$ then 
         $\SF{Badness}(x) \leq \Min\Sub{q \in P'}\SF{Badness}(q)$, which
         means that $\SF{Badness}(x) < \SF{Badness}(r)$. 
         From this finally follows: 
            $\SF{Badness}(r) > \Max\Sub{q \in R'}\SF{Badness}(q)$.
   \end{enumerate}
\end{proof}
This property implies that the result (the $R$) of a point to point query
is returned in monotonous increasing steps. 
Which means that when presenting the results to the user, the list box can
be filled in incremental steps by repeatedly selecting the \SF{MORE} option.

The increment operation, as such, can not yet be used to calculate the best 
solutions which are presented in the listboxes as depicted in 
\SRef{\PPQComplete}. 
For this latter purpose the $\SF{List}\Sub{f,t}(P,S)$ operation is 
introduced, which serves as a `driver' function for the entire process. 
\[ \begin{array}{l}
   \SF{List}\Sub{f,t}:\Powerset(\SF{Path}) \Carth \Powerset(\SF{Path}) \Carth \Powerset(\SF{Path}) 
      \Func \Powerset(\SF{Path}) \Carth \Powerset(\SF{Path})\Eol
   \SF{List}\Sub{f,t}(P,S,R) ~\Eq~ 
      \left\{ \begin{array}{ll}
         \tuple{P', S', R'}             & \mbox{if~}      R' \not= R\Eol
         \SF{List}\Sub{f,t}(P', S', R') & \mbox{else if~} P  \not= \emptyset\Eol 
         \Undefined	                & \mbox{otherwise}
      \end{array} \right.
\end{array} \]
where $\tuple{P',S',R'} =  \SF{Increment}\Sub{f,t}(P,S)$

This function calls the increment operation until it has come up with some new solutions ($R' \not= R$) 
or the pool of potential solutions has been exhausted ($P = \emptyset$). To provide the initial filling 
for the listboxes of a point to point query from $f$ to $t$, this function should be applied as: 
\[ \tuple{P,S,R} = \SF{List}\Sub{f,t}(\setje{[f]},\emptyset,\emptyset) \]
Now $R$ contains the set of found paths to be listed in the listbox. If users desire to see more options they 
can select the \SF{MORE} option (see \SRef{\PPQComplete}). This results in another call of the
$\SF{List}\Sub{f,t}$ function using $P$, $S$, $R$ as parameters.

Finally, the paths resulting from the search through the graph need to be translated into 
linear path expressions. For more details on linear path expressions please refer to 
\cite{Report:91:Hofstede:LISA-D}. 
In a later stage, however, the current definition of the linear path expressions as provided 
in \cite{Report:91:Hofstede:LISA-D} needs to be changed to better match our requirements. 
Every (proper) path through the graph can be translated into a linear path expression by the 
following recursive function:
\[ \begin{array}{l}
   \SF{PathExpr}: \SF{Path} \Func \SF{PathExpressions} \Eol
   \SF{PathExpr}([x_0,l_1,x_1,\ldots,l_n,x_n]) ~\Eq~
      x_0~\SF{Connector}(l_1,x_1)~x_1~\ldots~\SF{Connector}(l_n,x_n) 
\end{array} \]
where 
\[ 
   \SF{Connector}(l,x) ~\Eq~
      \left\{ \begin{array}{ll}
        \Conc               & \mbox{if~} l \in \setje{\Poly,\Spec}\Eol
        \Conc l       \Conc & \mbox{if~} x \in \RelTypes \land l \in \Roles(x)\Eol
        \Conc \Rev{l} \Conc & \mbox{otherwise}
   \end{array} \right.
\] 
Note that when $n=0$, we have: $\SF{PathExpr}([x_0]) = x_0$.

The linear path expressions are for internal use only. 
They can be mapped to proper SQL queries on the one hand, and verbalised as 
semi-natural language sentences using the verbalisation information as 
provided in the conceptual schema on the other hand. 
As stated before, the verbalisation of path expressions is the subject of 
further research.
 
   \section{Optimisation by Pre-Compilation}
\SLabel{section}{Reduce}

When humans look at a conceptual schema to find a path between two object 
types, they are usually able to identify parts of the schema which can safely 
be ignored when a searching for the actual path. 
In \SRef{\SubSchemas} such a situation is illustrated. 
In this schema, $F$ is the starting point of the point to point query and 
$T$ the end point. 
The three clouds represent subschemes. 
It is clear that subschema III can be safely ignored when searching for a 
path from $F$ to $T$ since the only way in/out of subschema III is through 
object type $B$. 
If a path would enter subschema III through $B$ (either continuing via fact 
type $f$ or $g$), the path would not be able to leave the subschema without 
creating a cycle. 
Such situations are not rare for real life applications. 
For instance, \cite{Book:94:Halpin:ORM} contains quite a number of schemes of 
real life applications with a similar pattern.

In this section a strategy is developed that allows us to reduce the graph 
associated with a conceptual schema before actually commencing a search. 
A possible way to approach this is to define a pruning algorithm that 
repeatedly cuts of irrelevant leaves from the graph. 
However, in the situation sketched in \SRef{\SubSchemas}, subschema III 
contains a cycle making it impossible for such a simple pruning algorithm to 
remove the entire subschema III. 
In this section we therefore develop a strategy for the removal of parts of 
the graph that may contain cycles. 
First a clustering algorithm is developed. 
After this, we use the simple pruning algorithm to remove irrelevant leaves, 
resulting in the removal of irrelevant subschemas even when they contain 
cycles (for instance subschema III).
{\def\Scale{0.55} \EpsfFig[\SubSchemas]{Connected subschemas}}

\subsection{Clustering a Graph}
The first step in our approach is the clustering itself. Clustering can be done by a pre-compilation of the 
conceptual schema. This pre-compilation should be done after the conceptual schema has been finished, 
but before the users start formulating queries. Although a conceptual schema is expected to evolve in the 
course of time (\cite{Report:92:Falkenberg:EISPlea}, \cite{PhdThesis:94:Proper:EvolvConcModels}, 
\cite{Report:92:Falkenberg:EISPromo}), the pre-compilation we propose here will not be costly to do after 
each evolution step. For small schemes one might even consider combining the pre-compilation with the 
search though the conceptual schema itself.

Formally, a clustering of a graph can be modelled as a function: $C: \N \PartFunc \Powerset(\SF{Node})$. 
An existing cluster $i$ within an existing clustering $C$ can be extended 
with nodes $E$, using the $C \bigoplus_i E$ operation. 
This operation is identified by:
\[ 
   (C \oplus_i E)(j) ~\Eq~ 
      \SF{if~} i=j \SF{~then~} C(j) \union E \SF{~else~} C( j ) \SF{~fi}
   \mbox{~~~~for each~} j \in \N 
\]
In the returned clustering the existing cluster $C(i)$ is grown to 
$C(i) \union E$.
 
A clustered graph can in itself be seen as a graph. 
The nodes are the clusters (effectively subgraphs of the original graph), and the edges can be derived from 
the original graph by having an edge between nodes (clusters) in the hypergraph if nodes in the clusters 
are connected in the original graph. This corresponds to the notion of a hypergraph since the clusters are 
treated as nodes. Obviously, the hypergraph can also be clustered, leading to yet another hypergraph. In 
the algorithms discussed here, we will repeatedly make use of the hypergraph notion. The idea is to use 
the hypergraphs to repeatedly simplify the graph, allowing us to identify irrelevant subschemas (which 
will correspond to nodes in one of the hypergraphs).

An important concept when clustering is the degree of a node. Let $G.E$ be the set of edges in a graph, 
then we define the degree of a node $n$ within the context of a set of nodes $N$ to be the number of nodes 
in $N$ that can be reached from $n$ by an edge in $G.E$. The word {\it reached} should be interpreted 
here as either directly connected, or one of the nodes contained in an involved cluster is connected. This 
notion of degree can be captured formally as:
\[ \begin{array}{l}
   \SF{Deg}: \Powerset(\SF{Node}) \Carth \SF{Node} \Func \N\Eol
   \SF{Deg}(N,n) ~\Eq~ 
     \card{\Set{ m \in N \SetMinus \setje{n}}{m \leftrightsquigarrow n}}
\end{array} \]
The principle of nodes (which could actually be clusters) being reachable from other nodes is represented 
by the $\leftrightsquigarrow$ operator:
\[ 
   n \leftrightsquigarrow m  \iff  
   \Ex{x,y}{~\setje{x,y} \in \Proj_1(G.E) \land x \prec n \land y \prec m~} 
\]
The expression $x \prec y$ captures the intuition of a node $x$ being equal to 
node $y$ or node $x$ being contained in the cluster $y$ (note that we will later on
use hyper clustering, resulting in nodes which contain other nodes). 
The $\prec$ operator is therefore defined by the following recursive rule:
\[ n \prec m  \iff  n = m  \lor  \Ex{m' \in m}{n \in m'} \]
From the above defined notion of degree we derive the so called {\it normalised} degree of a node. The 
idea behind the normalised degree is that the leaves of a graph (nodes with $\SF{Deg}(N,n) = 1$) can 
safely be ignored by the clustering algorithm as they will never lead to a cycle. For any node with a higher 
degree, a closer investigation is required, i.e. the clustering algorithm needs to be applied. 
Informally, the normalised degree is the number of neighbouring nodes with a 
degree higher then 1.
The formal definition of the normalised degree is:
\[ \begin{array}{l}
   \SF{NDeg}: \Powerset(\SF{Node}) \Carth \SF{Node} \Func \N\Eol
   \SF{NDeg}(N,n) ~\Eq~ 
      \card{\Set{m \in N \SetMinus \setje{n}}{
                 \SF{Deg}(N,m) > 1 \land m \leftrightsquigarrow n}} 
\end{array} \]
As an example, consider \SRef{\NDegEx}. There a graph is depicted where each node is labelled with the 
\SF{NDeg} of the node.

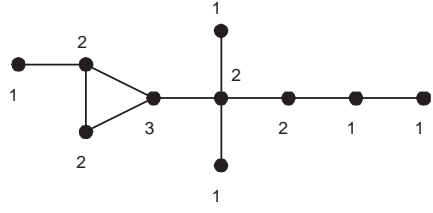
\EpsfFig[\NDegEx]{Normalised degrees of nodes.}
We now have enough primitives to define the clustering algorithm itself. The clustering needs to be done 
in such a way that the clusters themselves contain no branches (nodes $n$ with $\SF{NDeg}(n) > 2$). For 
clustering three functions are introduced. The first function is simply used to provide a nice interface to 
the other two clustering functions:
\[ \begin{array}{l}
   \SF{Cluster}: \Powerset(\SF{Node}) \Func (\N \PartFunc \Powerset(\SF{Node})) \Eol
   \SF{Cluster}(N) ~\Eq~ \SF{DoCluster}(\tuple{N,C},1)
\end{array} \]
where $C$ is the initial clustering defined as $C(i) = \emptyset$ for $1 \leq i \leq \card{N}$.

To cluster a graph $G$, this function should be invoked as: $\SF{Cluster}(G.N)$. The second clustering 
function (\SF{DoCluster}) is the `driver' function of the clustering algorithm, and the third function 
(\SF{Propagate}) the `propagation' function. The driver function takes three parameters. The first 
parameter represents the set of nodes that have not been placed in any cluster yet. The second parameter is 
the current clustering, and the last parameter is the number of the cluster that is currently being formed. 
This function selects, if there is any node left to be clustered, a node with the minimal degree and forms a 
cluster for this node (using the specified cluster number). Each time a new cluster is formed, the 
`propagation' function \SF{Propagate} is called, which will then try to extend the newly formed cluster. 
The driver function is identified by:
\[ \begin{array}{l}
   \SF{DoCluster}: 
         (\Powerset(\SF{Node}) \Carth (\N \PartFunc \Powerset(\SF{Node}))) 
         \Carth \N 
         \Func (\N \PartFunc \Powerset(\SF{Node}))\Eol
   \SF{DoCluster}(\tuple{N,C},i) ~\Eq~
      \left\{ \begin{array}{ll}
         C	& \mbox{if~} N = \emptyset\Eol
         \SF{DoCluster}(
            \SF{Propagate}(N \SetMinus \setje{n}, C \oplus_i \setje{n}, i),
            i+1
          )	& \mbox{otherwise}
      \end{array} \right.
\end{array} \]
where $n \in N$ such that 
   $\SF{NDeg}(N,n) \leq \Min\Sub{m \in N}\SF{NDeg}(N,m)$.

The propagation function tries to extend the size of the cluster. 
It does so by trying to find unclustered nodes which have a normalised degree
that is less then or equal two, and are connected to a node which is already 
contained in the new cluster. 
Note that in the determination of the normalised degree for the unclustered 
nodes, the nodes that are already contained in the new cluster are still
treated as unclustered nodes.
The clause `less than or equal to two' is absolutely essential to maintain the
idea of a cluster not (directly) containing any branches. 
Doing this allows us to safely prune the hypergraph, i.e. cutting away 
irrelevant parts. 
Furthermore, any simple leaves ($\SF{Deg}(N,n) = 1$) connected to a node in the 
new cluster become part of the cluster as well. 
The definition of the $\SF{Propagate}$ function is now provided by:
\[ \begin{array}{l}
   \SF{Propagate}: \Powerset(\SF{Node}) \Carth 
                   (\N \PartFunc \Powerset(\SF{Node})) \Carth \N \Func 
                   \Powerset(\SF{Node}) \Carth 
                   (\N \PartFunc \Powerset(\SF{Node}))\Eol
   \SF{Propagate}(N,C,I) ~\Eq~
      \left\{ \begin{array}{ll}
         \SF{Propagate}(N \SetMinus I, C \oplus_i I, i) & \mbox{if~} I \neq \emptyset\Eol
         \tuple{N,C}                                    & \mbox{otherwise}
      \end{array} \right.
\end{array} \]
where 	
\begin{eqnarray*}
   I &=&      \Set{n \in N - C(i)}{
                 \Ex{n' \in C(i)}{~\SF{NDeg}(N \union C(i), n') \leq 2 
                                   \land n \leftrightsquigarrow n'}~}\Eol
     &\union& \Set{n \in N - C(i)}{
                 \Ex{n' \in C(i)}{~\SF{Deg}(N \union C(i), n') = 1 
                                   \land n \leftrightsquigarrow n'}~}
\end{eqnarray*} 

\EpsfFig[\ClusterExA]{Example Clustering}
As an example consider the graph depicted in \SRef{\ClusterExA}. Each node has associated the number 
of the cluster it is part of. The arrows indicate the start node of each cluster. 
\begin{description}
   \item [Cluster 1:]
     At the start of the algorithm four nodes have an \SF{NDeg} of 1:
     the two right most nodes of cluster 1, and the top and bottom nodes of
     cluster 3.
     We arbritrarily chose for the first node of cluster 1 as our starting 
     point.

     The second node of cluster one is added because it has a neighbour 
     (in cluster one) with an \SF{NDeg} of 1. 
     The third node (the most left one) of cluster 1 is then added to 
     cluster one as it also has a neighbour (in cluster 1) with an \SF{NDeg}
     of 1 (the middle node of cluster 1).
 
     After adding the third node to cluster one, no more nodes can be added 
     since the third node has an \SF{NDeg} of 4. 
     Now that cluster one cannot be extended any further a new cluster is 
     formed.
 
   \item [Cluster 2:]
     At this moment, again four nodes have an \SF{NDeg} of 3: the two right 
     most nodes of cluster 2, and the top and bottom nodes of cluster 3 
     (remember, the nodes from cluster 1 have now been `removed' from the 
      \SF{NDeg} count). 
     
     Again we arbritrarily select the most right node of cluster 2 as a 
     starting point. 
     The three other nodes of cluster 2 are then added consecutively from
     right to left since they each have neighbours (in cluster 2) with an 
     \SF{NDeg} less than or equal to 2.

     The last node added to cluster 2 has an \SF{NDeg} of 4 and therefore no 
     further nodes can be added.

   \item [Cluster 3:]
     After the completion of cluster 2, once again four nodes with an 
     \SF{NDeg} of 3 remain.
     They are the four right most nodes of cluster 3.
     Note that the middle node of cluster 3 has an \SF{NDeg} of 1 as well since
     it has only one neigbour with a degree higher than 1.

     One arbritrary node is selected, and the other three nodes on cluster 3
     are added consecutively.

   \item[Cluster 4, 5 and 6:]
     All nodes that remain are the nodes of clusters 4, 5 and 6. 
     All three clusters are simple cycles of three or six nodes.
     After selecting one node on each of the cycles, the remaining nodes
     on the cycles are added to the clusters.
\end{description}
For the clustering algorithm we can prove some useful properties. Firstly, the clustering algorithm leads to 
a partition of the nodes in the graph:
\begin{lemma}
   For every graph $G$ the function $\SF{Cluster}(G.N)$ results in a partition of the nodes in 
   graph $G$.
\end{lemma}
\begin{proof}
   This follows immediately from the following observations:
   \begin{enumerate}
      \item The clustering algorithm only stops when all nodes are clustered (the $N = \emptyset$ clause 
         in the definition of $\SF{DoCluster}$).
      \item The clustering algorithm removes every clustered nodes from the `to do' set 
         (the $N \SetMinus \setje{n}$, and $N - I$ clauses in the definition of 
          $\SF{DoCluster}$ and $\SF{Propagate}$ respectively)
   \end{enumerate}
\end{proof}

The clusters do not contain nodes with a normalised degree that is higher than 2.
\begin{lemma}
    If $C = \SF{Cluster}(G.N)$, then for all $i \in \SF{Dom}(C)$:
      \[ \Al{n \in C(i)}{\SF{NDeg}(C(i),n) \leq 2 } \]
\end{lemma}
\begin{proof}
   Follows directly from the definition of $I$ in the definition of $\SF{Propagate}$.
\end{proof}
The consequence of this lemma is, as stated before, that there are no real decision points (branches) within 
one cluster. A node may have more then one leaf neighbour, but will not contain more then two non 
leaf neighbours. The result of this is that we can safely treat the clusters of one (hyper) graph as nodes on 
the next hyper graph, and remove them if they are found irrelevant for the point to point query. The 
different levels of hypergraphs are built using the hyper cluster function which continues clustering until 
a graph results that does not contain any cycles:
\[ \begin{array}{l}
   \SF{HCluster}: \Powerset(\SF{Node}) \Carth \N \Func \Powerset(\SF{Node}) \Carth \N\Eol
   \SF{HCluster}(N,i) ~\Eq~
      \left\{ \begin{array}{ll}
         \tuple{N,i}	          & \mbox{if~} \card{E} = \card{N} \SetMinus 1\Eol
         \SF{HCluster}(N', i + 1) & \mbox{otherwise}
      \end{array} \right.
\end{array} \]
where
\begin{eqnarray*}
   N' &=& \FuncRan(\SF{Cluster}(N))\\
   E  &=& \Set{\setje{x,y} \subseteq N }{x \neq y \land x \leftrightsquigarrow y }
\end{eqnarray*}
Note that $\FuncRan(f)$ returns the range of function $f$.
Not furthermore that a connected graph $G$ is acyclic exactly when 
$\card{G.E} = \card{G.N} - 1$. 
The initial call of the hyper cluster function is $\SF{HCluster}(G.N,1)$.

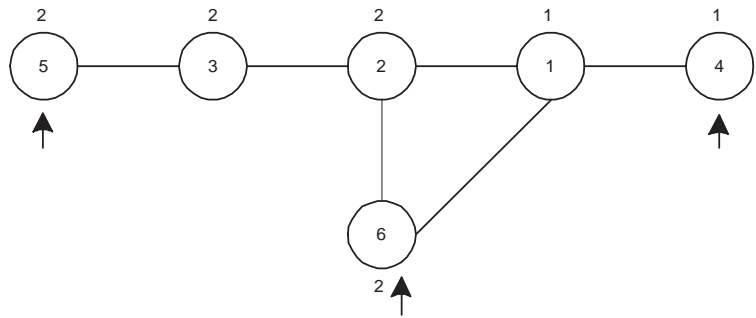
\EpsfFig[\ClusterExB]{First Level Hypergraph}
In \SRef{\ClusterExB} the hypergraph that can be associated to the clustering 
in \SRef{\ClusterExA} is depicted, together with a clustering of the hypergraph. 
Based on the clustering of this hypergraph, a second level hypergraph as 
depicted in \SRef{\ClusterExC} can be derived. 
As this graph is acyclic, the \SF{HCluster} function terminates after the 
completion of this clustering.
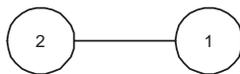
\EpsfFig[\ClusterExC]{Second Level Hypergraph}

\subsection{The Complexity of Clustering}
An important aspect of the clustering algorithm is the complexity of both the storage of the clusters as 
well as the algorithm itself. One call of the Cluster function is clearly linear in terms of the total number 
of nodes in the graph: $\Omega(\card{G.N})$. The calculation of the complexity of the $\SF{HCluster}$ 
function, however, is a bit more complicated. The $\SF{HCluster}$ repeatedly tries to reduce the number 
of nodes in the (hyper) graph by calling the $\SF{Cluster}$ function. Trying to get a grips on the 
complexity of the $\SF{HCluster}$ function thus requires us to analyse the expected number of times that 
the $\SF{Cluster}$ function will be called, i.e. how many levels of hypergraphs we expect to have. 

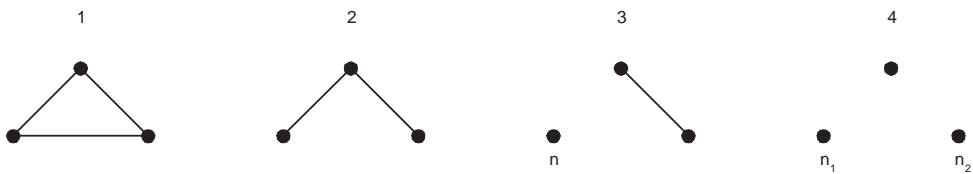
\EpsfFig[\RemainingNodes]{Alternative situations for remaining nodes}
A first observation we make is that we are always dealing with a connected graph since a conceptual 
schema is a connected graph. We now prove that performing \SF{Cluster} on a connected (hyper) graph 
with $n > 2$ nodes always leads to a connected hypergraph with maximally $n - 2$ nodes.
\begin{lemma} 
   $C = \SF{Cluster}(N) \implies \card{\SF{Dom}(C)} \leq \card{N} - 2$,  
   i.e. applying \SF{Cluster} leads to a reduction in the number of nodes of at least 2
\end{lemma}
\begin{proof}
   If $\card{N} = 3$, the $\SF{Cluster}(N)$ graph only contains 1 node, implying a reduction of exactly 
   two nodes. This follows directly from the fact that in a graph with three nodes $\SF{NDeg}$ has a 
   maximum value of 2.

   Let $\card{N} > 3$. The $\SF{Cluster}$ function does not terminate until all nodes have been 
   clustered. Now let us consider the last three nodes in $N$ that are selected by $\SF{Cluster}$ to be 
   clustered last. In \SRef{\RemainingNodes}, the four possible ways in which these three nodes can be 
   connected are depicted. In the first two cases, the three nodes would be clustered together, thus 
   leading to a reduction of at least two nodes. The two remaining cases require a closer examination:
   \begin{enumerate}
      \item In case three, the remaining three nodes obviously result in two clusters; leading to a reduction
         of only one node. However, we can prove that there must exist another cluster with at least two
         elements, thus implying that the total reduction size is still at least two.

         As the original graph is a connected one, node $n$ must have been connected to some node(s) which
         are already clustered. Let $M$ be the set of node(s) connected to node n that have been clustered the
         last. So the nodes in $M$ are all part of the same cluster, and have the highest cluster number of
         $n$'s neighbours. If $M$ contains more then one node, this consequently means that there is at least
         one other cluster with more than two nodes, thus leading to a reduction of at least two nodes.
 
         If $M$ contains only one node, say $m$, this implies that at the moment that the cluster containing
         node $m$ was formed, node $n$ was only connected to $m$. So node $n$ is a leaf node of the
         graph at that moment with an $\SF{NDeg}$ of 1. This means that node $n$ should have been
         clustered in the same cluster as $m$, which implies that $M$ contains at least two nodes.

      \item In case four, the remaining three nodes lead to three separate clusters. However, we will prove
         that there are enough clusters with more then one element to ensure a reduction of 2 nodes.

         Let $M_1$ be the set of node(s) connected to $n_1$ that were clustered last, and similarly $M_2$ the
         set of node(s) connected to $n_2$ that were clustered last. As the original graph was a connected
         graph, such nodes must exist.

         If $M_1$ or $M_2$ contains only one node, then they should have been clustered already (same
         argument as above). So $M_1$ and $M_2$ both contain at least two nodes. This is not yet enough,
         since $M_1$ and $M_2$ may overlap. However, for $i \in \setje{1,2}$ we have the following:
         \begin{quote}
            If $M_i$  has two elements, $n_i$  must have an $\SF{NDeg}$ of at most 2. This means that
            one of the two nodes in $M_j$ must have an $\SF{NDeg}$ of at most 2, since the clustering
            always starts with a node with the minimal $\SF{NDeg}$. This in turn means that $n_i $ is
            connected to a node with an $\SF{NDeg}$ that is less then or equal to two, and should therefore
            have been in the same cluster as $M_i.$.
         \end{quote}  
         As a result, $M_1$ and $M_2$ contain at least three nodes. So even if they overlap the reduction
         of two nodes is still guaranteed.
    \end{enumerate}
\end{proof}
The connectedness of a hypergraph after applying \SF{Cluster} to a connected graph follows directly from 
the way in which the edges are derived from the original graph. The above lemma allows us to identify the 
(execution) complexity of the \SF{HCluster} algorithm. Since the number of nodes in the graph decreases 
by two in every clustering of the \SF{HCluster} algorithm, the number of calls of \SF{HCluster} to 
\SF{Cluster} is limited to: $\lceil(\card{G.N})/2\rceil$. 
Therefore, the total complexity of the \SF{HCluster} function is: 
   $\Omega(\card{G.N}^2)$. 
However, since most conceptual schemes result in a sparse graph (not 
containing many edges), the results are likely to be better for most schemes. 

Another important issue is the complexity of the memory used. Every clustering requires the storage of the 
nodes in the cluster. Let $n = \card{G.N}$, and $k = \lfloor n/2 \rfloor$, then we have the following worst 
case with respect to the number of nodes that need to be stored:
\begin{eqnarray*}
   \Sigma_{j=0}^{k-1}   (n - 2i) + 1 
   &=& \Sigma_{i=1}^{k} (n + 2 - 2i) + 1\\
   &=& nk + 2k + 1 - 2 \Sigma_{i=1}^{k}i\\
   &=& nk + 2k + 1 - 2 \frac{k(k+1}{2}\\
   &=& nk + 2k + 1 - k(k + 1)\\
   &=& nk - k^2 + k + 1\\
   &=& \frac{n^2}{2} - \frac{n^2}{4} + \frac{n}{2} + 1\\
   &=& \frac{n^2}{4} + \frac{n}{2}   + 1
\end{eqnarray*}
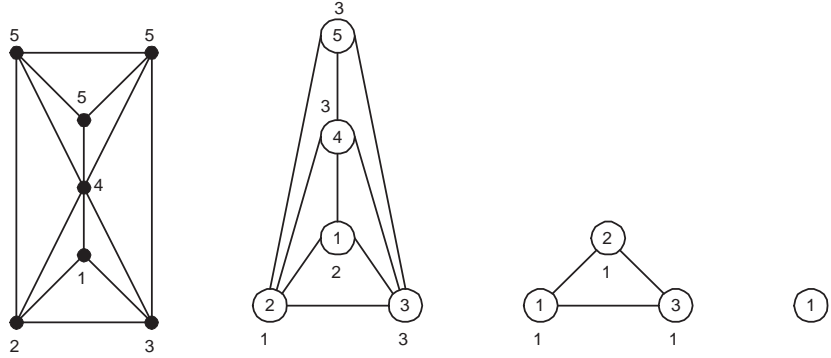
\EpsfFig[\WorstCase]{Worst Case Clustering}
As an example of a worst case graph, consider \SRef{\WorstCase}. 
This figure depicts the original graph, and the three associated hypergraphs 
after subsequent clustering steps. 
The original node contains 7 nodes, and it takes 3 steps to reduce it. 
The total number of nodes that needs to be stored is 16.

\subsection{Reducing the Search Space}
Using the hypergraphs, we can now reduce the size of the graph prior to searching the paths by means of 
the algorithm defined in \SRef{Search}. In general, a (hyper)graph is reduced by:
\[ \begin{array}{l}
   \SF{ReduceHG}\Sub{f,t}: \Powerset(\SF{Node}) \Func \Powerset(\SF{Node})\Eol
   \SF{ReduceHG}\Sub{f,t}(N) ~\Eq~
      \left\{ \begin{array}{ll}
         N                          & \mbox{if~} N=N'\Eol
         \SF{ReduceHG}\Sub{f,t}(N') & \mbox{otherwise}
      \end{array} \right.
\end{array} \]
where 	
\[ N' ~=~ \Set{ n \in N }{\SF{SDeg}(N,n) > 1 \lor t \prec n \lor f \prec n } \]
Note that, as mentioned before, the edges of the (hyper) graphs used in the above algorithm are derived 
from the original graph $G$. The {\it surrounding degree} (\SF{SDeg}) of a node $n$ in the (hyper) 
graph is the number of nodes in the original graph $G$ that are reachable from $n$, and that are not 
contained in (or equal to) the current node $n$. These nodes are the surroundings of node $n$. The formal 
definition therefore is:
\[ \begin{array}{l}
   \SF{SDeg}: \Powerset(\SF{Node}) \Carth \SF{Node} \Func \N\Eol
   \SF{SDeg}(N,n)  ~\Eq~ 
      \card{\Set{ y \in G.N}{
                 \Ex{m \in N \SetMinus \setje{n}}{
                     n \leftrightsquigarrow m \land y \prec m
                 }
            }}
\end{array} \]
\EpsfFig[\ExReductA]{Example Point to Point Query}

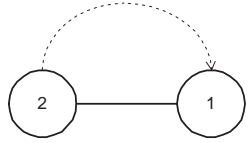
\EpsfFig[\ExReductB]{Point to Point Query on the Second Level Hyper Graph}
The reduction function $\SF{ReduceHG}\Sub{f,t}$ simply removes all nodes that 
are neither the start nor the end of the point to point query and have a 
surrounding of only one node (e.g. subschema III in \SRef{\SubSchemas}). 
The completely reduced graph is calculated by the following `driver' function:
\[ \begin{array}{l}
   \SF{Reduce}\Sub{f,t}: 
      \Powerset(\SF{Node}) \Carth \N \Func \Powerset(\SF{Edge}) \Carth \Powerset(\SF{Node})\Eol
   \SF{Reduce}\Sub{f,t}(N,n) ~\Eq~ 
      \left\{ \begin{array}{ll}
         \tuple{E',N'}                         & \mbox{if~} N=N'\Eol
         \SF{Reduce}\Sub{f,t}(\union N', n -1) & \mbox{otherwise}
      \end{array} \right.
\end{array} \]
where
\[
   N' ~=~ \SF{ReduceHG}\Sub{f,t}(N) \mbox{~and~}
   E' ~=~ \Set{\tuple{P,l} \in E}{P \subseteq N}
\] 
If $\SF{HCluster}(G.N,0) = \tuple{N',n}$ for a certain graph $G$, then the search should be performed 
in the reduced graph: $\tuple{E'',N''} = \SF{Reduce}\Sub{f,t}(N',n)$. The interesting question now is 
when $\SF{HCluster}(G.N,0)$ should be calculated. Since this latter call does not depend on the point to 
point query specific source and destination, it could be calculated once after the completion of the 
conceptual schema (from which graph $G$ is derived). Alternatively, one could calculate the hyper 
clustering each time a point to point query needs to be completed, which is likely to be very costly.

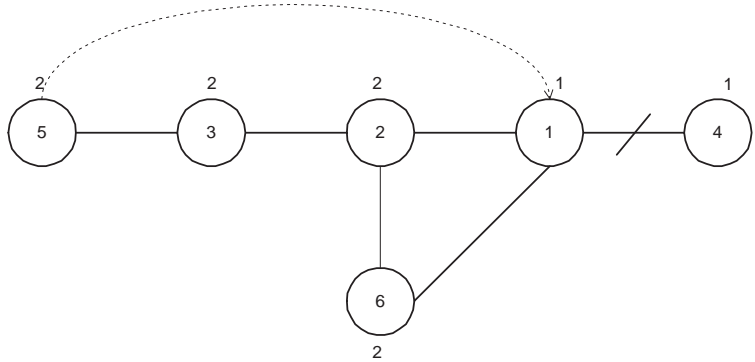
\EpsfFig[\ExReductC]{Point to Point Query on the First Level Hyper Graph}
As an example reduction, consider the point to point query denoted in \SRef{\ExReductA}. The reduction 
algorithm starts with the reduction of the top level hypergraph as depicted in \SRef{\ExReductB}. 
Obviously, this graph cannot be reduced in any way. The next hypergraph that is considered by the 
reduction algorithm is shown in \SRef{\ExReductC}. Cluster 3 (containing the large cycle) is connected to 
only one other node and does not contain the start or end of the point to point query. Therefore, it can 
safely be removed from the graph. Finally, the original search graph can be reduced; the resulting graph is 
shown in \SRef{\ExReductD}. In this graph, the two leaf nodes from cluster 4 can be removed, as well as the 
two right nodes of cluster 1. Note that the nodes from cluster 3 have already been removed as the entire 
cluster was already removed in the previous step of the Reduce function. 
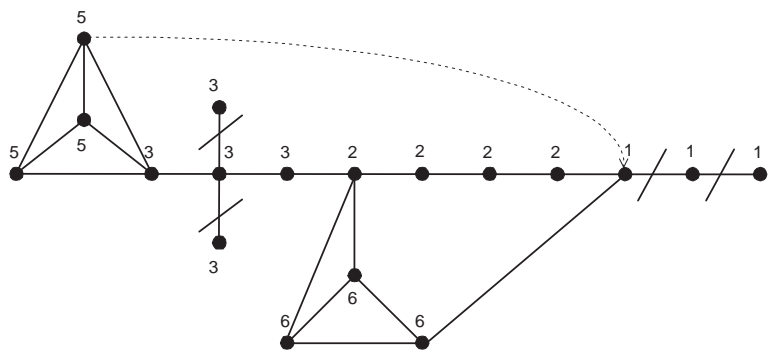
\EpsfFig[\ExReductD]{The Reduced Search Graph}

   \section{Conclusions}
\SLabel{section}{Concl}
In this article we introduced a novel way for computer supported query formulation called point to point 
queries. We provided a sample session with a provisional tool supporting point to point queries, and 
briefly discussed the relationship to query by navigation and query by construction. Together with these 
mechanisms a powerfull query formulation tool could be build. Furthermore, a search algorithm was 
introduced to search for the relevant paths between the specified points. Finally, an optimisation strategy 
for the search process was discussed based on a pre-compiled clustering of the conceptual schema graph.

As a next step, the path expressions should be further developed to suit our needs. Furthermore, elegant 
verbalisations of the path expressions should be catered for. 

   \AddBib{asy}
   \BIBLIOGRAPHY{alpha}
\end{document}